\documentclass[prb,aps,showpacs]{revtex4}

\usepackage{graphicx}
\usepackage{amssymb,amsmath}

        \def \o{\omega}    
\def \b{\beta}    \def \s{\sigma}     
\def \e{\epsilon}          
\def \d{\delta}        \def \l{\lambda}
    \def \o{\omega}

   \def \S{\Sigma}    
\def \D{\Delta}

\def \h{\hbar}   

\def \f{\frac}
\def \del{\partial}    

\def \ord{\mathcal{O}}

\def\lba{\left(}    \def\rba{\right)}
\def\lbc{\left[}    \def\rbc{\right]}

\def \bra{\langle}   \def \ket{\rangle}

\def\be{\begin{equation}}    \def\ee{\end{equation}}

\def \vp{{\bf p}}  
\def \vq{{\bf q}}  
\def \vk{{\bf k}}  

\def \tc{T_{\mathrm c}}   

\def \nbc{f_{\mathrm c}}	\def \c{{\mathrm c}}
\def \te{\tilde{\epsilon}}

\begin{document}

\title{Transition Temperature of Dilute, Weakly Repulsive Bose Gas}
\author{Masudul Haque}
\email{masud@physics.rutgers.edu}
\author{Andrei E. Ruckenstein}
\email{andreir@physics.rutgers.edu}

\affiliation{Department of Physics and Astronomy, Rutgers University,
136 Frelinghuysen Road, Piscataway, NJ-08854-0849, USA}

\date{\today}

%
%
\begin{abstract}

Within a quasiparticle framework, we reconsider the issue of computing
the Bose-Einstein condensation temperature ($T_{\rm c}$) in a weakly
non-ideal Bose gas.  The main result of this and previous
investigations is that $T_{\rm c}$ increases with the scattering
length $a$, with the leading dependence being either linear or
log-linear in $a$.  The calculation of $T_{\rm c}$ reduces to that of
computing the excitation spectrum near the transition.  We report two
approaches to regularizing the infrared divergence at the transition
point.  One leads to a $a\sqrt{|\ln{a}|}$-like shift in $T_{\rm c}$,
and the other allows numerical calculations for the shift.


\end{abstract}

\maketitle

\section{Introduction}   \label{sect_intro}

Inter-particle correlations in the weakly repulsive dilute Bose gas
have been studied for more than half a century.  It is therefore quite
remarkable that the dependence of the Bose-Einstein condensation
temperature ($\tc$) on the interaction strength has continued to
generate controversy until this time.  In this paper we present an
analysis of the longstanding $\tc$ problem starting from a
quasiparticle picture of the system in the un-condensed
(high-temperature) side.

The model is described by the Hamiltonian
\[
\hat{H} = \sum_{\vk} \e_{\vk} \hat{b}_{\vk}^\dag \hat{b}_{\vk}
+  \frac{1}{2V}\sum_{\vp,\vq,\vk} U(\vk)\, \hat{b}_{\vp+\vk}^\dag
\hat{b}_{\vq-\vk}^\dag \hat{b}_{\vp} \hat{b}_{\vq} \, \, ,
\] 
where $\e_{\vk}=k^2/2m$ is the free-gas spectrum, and $\hat{b}$,
$\hat{b}^\dag$ are bosonic operators.  In principle, one is interested
in any possible interaction function $U(\vk)$, but the simplest
possible form is a delta function in real space, so that $U(\vk) = U$
is momentum-independent.  To first order $U$ is related to the
$s$-wave scattering length $a$ by $U = 4\pi\h^2a/m$.  A dimensionless
measure of the interaction is the quantity $an^{1/3}$, where $n = N/V$
is the density.  The program is to examine the shift in the critical
temperature $\tc$, as compared to a noninteracting gas of the same
density, as a function of $an^{1/3}$, for small $an^{1/3}$.

For several decades, the quasiparticle picture has been a central
paradigm for studying correlated many-body systems.  Even in a highly
correlated Bose liquid like liquid helium, a quasiparticle description
accounts for the transition temperature extremely well; e.g., the
lambda point temperature can be obtained from the ideal gas expression
$\tc^{(0)} = \lba{2}\pi/m[\zeta(\tfrac{3}{2})]^{2/3}\rba n^{2/3}$ by
replacing the helium atomic mass $m$ by a quasiparticle effective
mass, $m$*.  The effective mass in the liquid exceeds the bare mass,
because of the inertia of the medium which has to make way for any
atom to move.  Correspondingly, the transition temperature is lower
for liquid $^4$He than would be the case for an ideal bose gas with
the same bare mass.

The question we would like to address is the following: how much
information can one get about $\tc$ from a quasiparticle description
of the weakly interacting Bose gas?  The simplest treatment, a Hartree
correction, does not give any contribution to the transition
temperature.  The exchange contribution at mean field level (Fock),
does not have any effect on $\tc$ for a momentum-independent
interaction, but gives a negative shift for a momentum-dependent
interaction \cite{fetter}.  The Fock contribution to $\tc$ depends on
the details of the interaction, and for small interactions is weaker
than the leading shift we are examining.  The present understanding is
that there is no mean field effect on the leading shift, and that
higher-order correlations are required to calculate this leading shift
in $\tc$.

Interaction effects in the non-ideal Bose gas can be described in
terms of the two-body $t$-matrix, i.e., the vacuum scattering
amplitude.  This quantity describes all possible collisions between
\emph{two} particles, without taking into account the fact that the
surrounding medium has an effect on these collisions.  Many-body
corrections arising from the surrounding gas can be treated by using
the many-body $T$-matrix, which is the sum of ladder diagrams.

In this paper, we approach the calculation of $\tc$ for a Bose gas
using perturbation theory in the two-body $t$-matrix.  This is to be
regarded as a first step toward a calculation using the many-body
$T$-matrix.  The idea is that, if one can regard the system as a
weakly interacting gas of quasiparticles, then perturbation theory in
the quasiparticle interactions should give a reasonable starting
point, and the effects of the surrounding medium can be treated as a
correction to these results.

The work in this paper is not meant to be a definitive determination
of $\tc$ for the weakly interacting bose gas, but instead pursues the
more limited aim of calculating $\tc$ within a quasiparticle
framework.  As is typical for descriptions of critical-point
properties, our approach faces infrared divergences.  We elaborate on
two ways to regularize these divergences.

During the past several years, work by Baym, Lalo\"e and collaborators
\cite{baym_prl,baym_N,baym_a2log,mueller_baym_finite,baym_bigpaper,
baymfriends_numerics}, has predicted a positive shift in $\tc$ which,
to leading order, is \emph{linear} in the scattering length; $\D\tc =
{c_1}an^{1/3} + \ord(a^2n^{2/3})$.  Attempts to calculate the
coefficient $c_1$, however, have continued to give fluctuating
results.  This has prompted the Baym group to predict
\cite{baym_a2log} a logarithmic contribution at the \emph{next} order,
i.e., a $-a^2\ln{a}$ contribution.  The presence of such a
non-analyticity might explain the difficulty in any numerical estimate
of $c_1$.

Stoof, from a renormalization-group analysis
\cite{stoof2,stoof_private}, has predicted an increase of the
transition temperature proportional to $a|\ln{a}|$, i.e., a
non-analyticity at \emph{leading} order.  A leading non-analytic
behavior would actually explain even better the widely varying results
obtained in attempts to calculate the coefficient $c_1$ based on the
assumption of linear shift $\D\tc \sim {c_1}an^{1/3}$.  One should
also note that the work of Baym and collaborators relies heavily on
power-counting and dimensional arguments. This kind of argument
typically does not catch logarithmic corrections.  The possibility of
leading non-analytic dependence of $\D\tc$ on the interaction,
therefore, needs to be further examined.

An older attack on the $\tc$ problem is that of Kanno
\cite{kanno12,kanno3}.  Using a ``quasi-linear'' canonical
transformation, Kanno calculated the free energy and hence $\tc$ of
the Bose gas.  Other derivations of the transition temperature include
the canonical-ensemble calculation \cite{wilkens} of Wilkens \emph{et
al} leading to a negative shift, Schakel's effective-action theory
\cite{schakel_ijmp,schakel_boulevard} leading to a prediction for a
positive linear shift, and Kleinert's recent five-loop calculation
\cite{kleinert}.  In addition, there have been spurious mean-field
level predictions \cite{huang_prl,toyoda} for shifts in $\tc$
proportional to the square root of the interaction.

One could think of studying the $\tc$ problem experimentally, e.g., in
the context of trapped atomic Bose gases.  Unfortunately, trap effects
dominate in the dependence of $\tc$ on the interaction for these
systems.  Size effects cause a decrease of $\tc$ at mean-field level,
far larger than the intrinsic effect under study in this paper.  The
$^4$He-Vycor system \cite{reppy} might be more suitable for studying
the $\tc$ problem, but is yet to reach the accuracies necessary for
identifying logarithm-like corrections.

\section{The Quasiparticle Approach}  \label{sect_qp}

The starting point of our analysis is the expansion of the single
particle Green function to leading nontrivial order in quasiparticle
interactions, namely:
\begin{multline} \label{greenfunction}
G(\vk, z) ~=~ \frac{1}{z-\lba\epsilon_\vk -\mu
+\Sigma_{\rm R}(\vk,E_\vk) \rba 
- \lba \Sigma (\vk,z) - \Sigma_{\rm R} (\vk,E_\vk) \rba}  \\
\approx~ \frac{1}{z-E_\vk} ~+~
\frac{[\Sigma(\vk,z)- \Sigma_{\rm R}(\vk, E_\vk)]}
{[z-E_\vk]^2} ~+~ \ldots  \,
= \frac{1}{z-E_\vk}  \lbc 1+\int \frac{d\omega}{\pi}
\frac{\Sigma_{\rm I}(\vk,\o)}{(\o -z)(\o -E_\vk)} \rbc
+ \ldots  \, \, .
\end{multline}
Here $\Sigma_{\rm R,I}(\vk,\o)= \mathfrak{Re},\mathfrak{Im}
\Sigma(\vk,\o +i0^+)$ are the real and imaginary parts of the self
energy, and $E_\vk = \e_\vk -\mu +\Sigma_{\rm R}(\vk,E_\vk)$ is the
quasiparticle energy.  This expression has a number of attractive
properties which would not appear in simple perturbation theory in the
inter-particle interactions.

In the quasi-particle approximation \eqref{greenfunction}, the
single-particle spectral function $\mathcal{A}(\vk,\o) \equiv -2
\mathfrak{Im} G({\vk},\o+i0^+)$ is given by:
\begin{equation}
\label{spectralfunction}
\mathcal{A}(\vk,\o) ~\approx ~ 2\pi \d(\o-E_\vk)
\lbc 1+\int \frac{d\o'}{\pi} 
\frac{\Sigma_{\rm I}(\vk,\o')}{(\o'-E_\vk)^2}\rbc 
- 2\frac{\Sigma_{\rm I} (\vk,\omega)}
{(\o'-E_\vk)^2}  \, \, .
\end{equation}
Note that the form \eqref{spectralfunction} automatically satisfies
the spectral sum-rule, $\int{d\o}\mathcal{A}(\vk,\o) =1$.  

We proceed by writing the particle number $N$ as a momentum sum over
the convolution of the spectral function and the Bose thermal
distribution function, $f(\o)=1/(e^{\b\o}-1)$.  
\begin{equation}  \label{N_qp_specfunc}
N ~=~ \sum_\vk \int_{-\infty}^{\infty} \frac{d\o}{2\pi}
\mathcal{A}(\vk,\omega) f(\o) ~=~ N_0 + V \int\frac{d^3 k}{(2\pi)^3} f(E_\vk) ~+~
V \int\frac{d^3 k}{(2\pi )^3} \int \frac{d\omega}{\pi}
\frac{[f(E_\vk)-f(\o)]}{(\o - E_\vk)^2} \Sigma_{\rm I} (\vk,\omega)  
\, \, .
\end{equation}

To proceed further we must resort to an explicit expression for the
self-energy.  We make the assumption that our quasi-particles interact
weakly, so that we can work perturbatively.  We limit ourselves to
terms up to second order in the interaction.   

\begin{figure}  
\includegraphics[width=6.0cm]{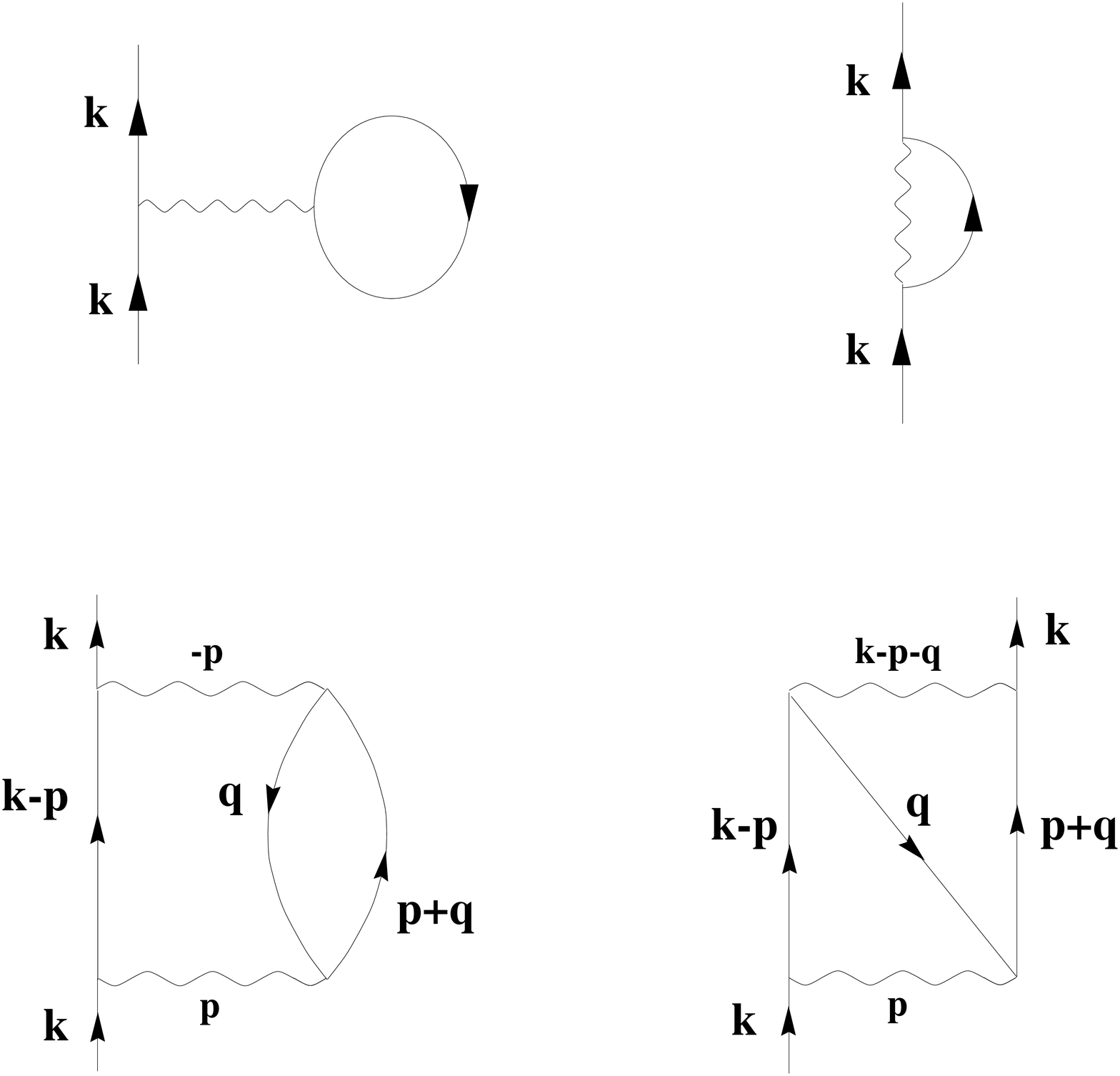}  
\caption{\label{fig_selfenergy12} The two first-order and two second-order
self-energy diagrams included in our calculation.  Of the six
topologically distinct second-order diagrams, only the two shown here
are necessary for momentum-independent interaction, because the other
four can be obtained by 1st-order self-energy insertions in the first 
order diagrams.}
\end{figure}

\subsection{Perturbation Theory}

We first perform perturbation theory for the self-energy in the bare
potential $U(\vk)$, retaining explicit momentum-dependence.  The
diagrams that need to be calculated are shown in figure
\ref{fig_selfenergy12}.  After performing the Matsubara sums, we get:
\begin{multline} \label{selfenergy}
\Sigma(\vk,z) ~\approx~  \int\frac{d^3 p}{(2\pi)^3}
\lbc U(\vk-\vp)+ U(0)\rbc f(\tilde{\e}_\vp) \\
+~  \int\frac{d^3 p}{(2\pi)^3} \int \frac{d^3 q}{(2\pi)^3} 
\frac{\lbc f(\tilde{\e}_\vq) 
+ f(\tilde{\e}_\vq)f(\tilde{\e}_{\vp+\vq}) 
+ f(\tilde{\e}_\vq)f(\tilde{\e}_{\vp-\vk}) 
- f(\tilde{\e}_{\vp+\vq})f(\tilde{\e}_{\vp-\vk}) \rbc}
{z- \lba \tilde{\e}_{\vp +\vq} +
\tilde{\e}_{\vp-\vk}-  \tilde{\e}_{\vq} \rba}
\times   \\
U(\vp) \lbc U(-\vp) + U(\vk-\vp-\vq)\rbc   \, \, .
\end{multline}
Here $\tilde{\e}_\vk = \e_\vk-\mu$.  For a momentum-independent
$U(\vk) =U$, the first-order Hartree and Fock terms give equal
contributions, as do the two second-order diagrams.  However, Using
$U(\vk) =U$ directly in \eqref{selfenergy} causes ultraviolet (UV)
problems in the second-order self-energy, due to the term with a
single Bose function.  The prescription for removal of such
divergences is to shift to a description in terms of the vacuum
scattering amplitude $t$, which is a more physical quantity than the
bare potential.  The bare potential $U(\vk)$ is related to the
$t$-matrix by
\begin{equation} \label{U_to_t}
U(\vk) = \mathfrak{Re}
\langle \vp-\vk,\vq+\vk\left|\hat{t}\right|\vp,\vq\rangle 
~-~ \int\frac{d^3k'}{(2\pi)^3} \frac{ \mathfrak{Re} \lbc
\bra\vp-\vk,\vq+\vk\left|\hat{t}\right|\vp-\vk',\vq+\vk'\ket
\bra\vp-\vk',\vq+\vk'\left|\hat{t}\right|\vp,\vq\ket \rbc}
{\e_\vp+\e_\vq-\e_{\vp-\vk}-\e_{\vq+\vk}}
\end{equation}
When eq \eqref{U_to_t} is used to replace the bare potential by the
$t$-matrix in eq \eqref{selfenergy}, we get an $\ord(t^2)$ term from
the first-order self-energy that exactly cancels the UV-divergent term
appearing at second order.  There is now no problem in using a
low-momentum limit in which the $t$-matrix reduces to a
momentum-independent constant $t = 4{\pi}\h^2a/m$.  The real
(on-shell) and imaginary parts of the self-energy now reduce to:
\begin{equation}   \label{SE_real}
\Sigma_{\rm R}(\vk,\tilde{\e}_\vk)  ~\approx~  2tn  
~+~ 2t^2\, \int\frac{d^3 p}{(2\pi)^3} \int\frac{d^3 q}{(2\pi)^3}
\frac{f(\te_{\vq})f(\te_{\vp+\vq}) +f(\te_{\vq})f(\te_{\vp-\vk})
-f(\te_{\vp+\vq})f(\te_{\vp-\vk})}
{\tilde{\e}_{\vq} +\tilde{\e}_{\vk}
-\tilde{\e}_{\vp+\vq} -\tilde{\e}_{\vp-\vk}}   \, \,  ,
\end{equation}
and
\begin{multline}   \label{SE_imag}
\Sigma_{\rm I}(\vk,\o)  \, \approx\   
- 2{\pi}t^2\, \int\frac{d^3 p}{(2\pi)^3} 
\int\frac{d^3 q}{(2\pi)^3}\ 
\delta \lba \o - \lbc \te_{\vp +\vq} + \te_{\vp-\vk} -\te_{\vq} \rbc
  \rba     \\
\times  \lbc f(\te_\vq) + f(\te_\vq)f(\te_{\vp+\vq}) 
+ f(\te_\vq)f(\te_{\vp-\vk}) - f(\te_{\vp+\vq})f(\te_{\vp-\vk})   
\rbc     \, \, .
\end{multline}

\subsection{Self-consistent Perturbation Theory}

We will use \eqref{SE_real} and \eqref{SE_imag} to define a
self-consistent quasi-particle approximation, formally second order in
the vacuum scattering amplitude.  This is done by replacing the
free-particle dispersion functions ($\e_\vq$'s) on the right hand
sides of \eqref{SE_real},\eqref{SE_imag} by the quasiparticle
dispersions, $E_\vq$'s.  The imaginary part now becomes
\begin{multline}
\S_{\rm I} (\vk,\o)  ~\approx~  
-2\pi t^2 \int \frac{d^3 p}{(2\pi)^3} \int\frac{d^3 q}{(2\pi)^3} \ 
\delta \lba \o - \lbc E_{\vp +\vq} + E_{\vp-\vk} -E_{\vq} \rbc
  \rba     \\
\times  \lbc f(E_\vq) + f(E_\vq)f(E_{\vp+\vq}) + f(E_\vq)f(E_{\vp-\vk}) 
- f(E_{\vp+\vq})f(E_{\vp-\vk})   \rbc     \, \, .
\end{multline}
When this expression for the imaginary part is used in eq
\eqref{N_qp_specfunc}, the $\S_{\rm I}$ contribution cancels out exactly,
resulting in
\begin{equation}  \label{n_qp_realonly}
n =   \int \frac{d^3 k}{(2\pi)^3} f \lba E_\vk\rba  \, \, ;
\end{equation}
i.e., the number of quasiparticles is equal to the number of
particles.  
Since we will be using these results above and at the transition, we
have neglected the zero-momentum occupancy $N_0$ as compared to
$N$.   
We note that the imaginary part of the self-energy, i.e., effects of
quasiparticle broadening, has dropped out of the expression for
particle number and hence does not play any role in our determination
of the transition temperature.  

Finally, making the real part of the self energy (eq \ref{SE_real})
self-consistent, we get the following expression for the
quasi-particle energy:
\begin{multline}    \label{E_qp_selfcons}
E_\vk = \e_\vk -\mu + \S_{\rm R}(\vk,E_\vk) \\
~=~  \e_\vk - \lba \mu -2tn \rba
+ 2t^2 \int \frac{d^3 p}{(2\pi)^3} \int\frac{d^3 q}{(2\pi)^3} 
\frac{f(E_\vq)f(E_{\vp+\vq}) + f(E_\vq)f(E_{\vp-\vk}) 
- f(E_{\vp+\vq})f(E_{\vp-\vk})}
{E_{\vq}+E_{\vk} -E_{\vp+\vq}-E_{\vp-\vk} } 
\, \, .
\end{multline}

\section{Critical Temperature and Spectrum}  \label{sect_tc+spectrum}

To derive an expression for the change in the critical temperature due
to quasi-particle interactions, we first consider the density of an
ideal Bose gas, both at its transition temperature $\tc^{(0)} =
\lba{2}\pi/m[\zeta(\tfrac{3}{2})]^{2/3}\rba n^{2/3}$, and at
temperature $\tc$ which is the transition point of the
\emph{interacting} Bose gas.
\begin{equation}  \label{n0_at_tc}
\int \frac{d^3k}{(2\pi)^3} 
\frac{1}{e^{\epsilon _{\bf{k}}/k_B \tc} -1} ~=~ n^{(0)}(\tc) 
~\approx ~
\lba\frac{\tc}{\tc^{(0)}}\rba^{3/2} n^{(0)}(\tc^{(0)})
~\approx~ \lba 1+ \frac{3}{2}\frac{\D\tc}{\tc^{(0)}} \rba 
n^{(0)}(\tc^{(0)}) \, \, .
\end{equation}
Here $\tc = \tc^{(0)} +\D\tc$.  We are looking for the difference of
transition temperatures of an interacting and a non-interacting gas at
the \emph{same density}, therefore $n^{(0)}(\tc^{(0)}) = n(\tc) = n$.
Evaluating \eqref{n_qp_realonly} at $\tc$ and subtracting equation
\eqref{n0_at_tc} we obtain the leading correction to $T_c$ due to
interactions:
\begin{equation}    \label{formula}
\frac{\D\tc}{\tc^{(0)}} ~\approx~
-\frac{2}{3n}  \int \frac{d^3  k }{(2\pi)^3} \ 
\lbc  f_\c (\xi_\vk) ~-~ f_\c (\e_\vk)  \rbc,
\end{equation}
Here $f_\c (x) =1/(e^{x/k\tc} -1)$ is the Bose distribution function
evaluated at the critical temperature; and $\xi_\vk = E_\vk(\tc) =
\e_\vk -\mu(\tc) +\S_{\bf R}(\vk,\xi_\vk) = \e_\vk +\S_{\bf
R}(\vk,\xi_\vk) -\S_{\bf R}(0,0)$ represents the single particle
excitation spectrum at $\tc$.  
The problem of calculating the relative shift has thus been reduced to
calculating the spectrum at the critical point, $\xi_\vk$.

At this stage, we are able to present a simple argument for the
\emph{increase} of $\tc$ due to the addition of a repulsive
interaction.  This is important in light of the continuing appearance
of negative-shift predictions in the literature \cite{toyoda,wilkens}.
The major contribution to eq \eqref{formula} comes from the
infrared; therefore approximately
\[
\D\tc \propto \int_0^{\rm cutoff} dk\ k^2 [\e_\vk^{-1}-\xi_\vk^{-1}] 
\, \, .  
\]
If the quasiparticle spectrum $\xi_\vk$ is ``harder'' than the bare
spectrum $\e_\vk = k^2/2m$, i.e., if $\xi_\vk$ is sub-quadratic in $k$
or quadratic with an effective mass $m^*<m$, then the $\e_\vk$
integral dominates and the shift is positive.  On the other hand if
the spectrum at $\tc$ were ``softened'' by the interaction, one would
have a decrease in the transition temperature.  Since the spectrum
$\xi_\vk$ at $\tc$ is known to be intermediate between linear and
quadratic, the weakly interacting Bose gas has a \emph{positive} shift
in the transition temperature with the introduction of a weak
repulsive interaction.

\subsection{Spectrum from Perturbation Theory}

It follows from \eqref{E_qp_selfcons} 
that $\xi_\vk$ satisfies the nonlinear integral equation,
\begin{multline}  \label{qpdisp}
\xi_\vk = \e_\vk 
+ 2t^2 \int \frac{d^3 p}{(2\pi)^3} \int\frac{d^3 q}{(2\pi)^3} \  
\frac{f(\xi_\vp)f(\xi_{\vq}) + f(\xi_\vp)f(\xi_{\vp+\vk-\vq}) 
- f(\xi_{\vq})f(\xi_{\vp+\vk-\vq})}
{\xi_{\vp}+\xi_{\vk} -\xi_{\vq}-\xi_{\vp+\vk-\vq} } 
\\
- 2t^2 \int \frac{d^3 p}{(2\pi)^3} \int\frac{d^3 q}{(2\pi)^3} \ 
\frac{f(\xi_\vp)f(\xi_{\vq}) + f(\xi_\vp)f(\xi_{\vp-\vq}) 
- f(\xi_{\vq})f(\xi_{\vp-\vq})}
{\xi_{\vp} -\xi_{\vq}-\xi_{\vp-\vq} } 
\end{multline}
In the perturbative quasiparticle approach, equations \eqref{formula}
and \eqref{qpdisp} completely determine $\D\tc$.

If the quasiparticle dispersion $\xi_\vk$ were quadratic at small
momenta, the self-energy integral in eq \eqref{qpdisp} would be
divergent in the infrared.  A power-counting of this equation shows
that, for self-consistency, the dispersion at small momenta should be
sub-quadratic, $\xi_\vk \sim k^{3/2}$, modulo possible logarithmic
factors which power-counting cannot predict.  The power-counting
arguments were first devised by Patashiskii and Pokrovskii
\cite{pata+pokrov}.

Unfortunately, this approach using second-order perturbation theory
over-modifies the infrared spectrum.  The long-wavelength spectrum at
the transition point should be $\xi_\vk \sim{k}^{2-\eta}$, where
$\eta$ is the anomalous dimension.  The present system falls in the
same universality class as the $N=2$ quantum rotor model.  For this
universality class, a recent Monte Carlo calculation
\cite{eta_montecarlo} has produced the value $\eta\approx{0.038}$.  In
contrast, eq \eqref{qpdisp} gives $\eta = 0.5$.

Acknowledging that the present framework over-modifies the spectrum at
$\tc$, we proceed to analyze eqs \eqref{formula} and \eqref{qpdisp}
in order to extract the shift in the critical temperature due to
interactions.

\subsection{Infrared Divergence}

There are several ways of dealing with the infrared divergence.  One
is to treat equation \eqref{qpdisp} self-consistently and determine the
self-consistent $\xi(\vk)$ numerically; this procedure is
outlined in section \ref{sect_selfcons}.  This approach has also been
explored by Baym and collaborators.

In the next section, we take a different approach, motivated by the
work of Kanno \cite{kanno3}.  We make an expansion in $\S_{\rm
R}(\vk,\xi_\vk)$ that allows us to use the chemical potential at the
critical point, $\mu(\tc) = \mu_\c$, as an infrared cutoff.  This
approach allows us to express the relative shift in the transition
temperature as an expansion in $\b_\c\mu_\c = \mu_\c/k_{\rm B}\tc$,
which itself can in turn be related to the scattering length.  This
leads to a transition temperature shift $\D\tc$ that is
$a\sqrt{\mathcal{K}(a)}$ to leading order, where $\mathcal{K}(x)$ is a
function approximately like $|\ln(x)|$, to be specified in the next
section.  The appropriateness of this procedure is difficult to
gauge, but it provides strong support to the possibility of
non-analytic corrections to the transition temperature at leading
order. 

Another infrared regularization procedure that has appeared in the
literature is A.M.J. Schakel's prescription of analytic continuation
\cite{schakel_boulevard}.  The connection between Schakel's treatment
and other calculations awaits further study.

\section{Non-SelfConsistent Calculation}  \label{sect_nonself}

\begin{figure}
\includegraphics[width=8.2cm]{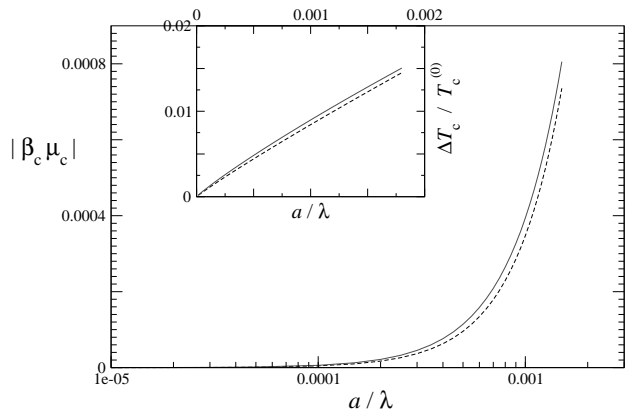}
\caption{ \label{fig_nonself_betamu} 
Approximation for eq \eqref{nonself_muc}.  Solid curve is exact
numeric solution; dashed curve shows approximation $|\b_\c\mu_\c|
\approx 16\pi(a/\l_\c)^2  |\ln(a/\l_\c)|$.  Inset shows comparison
for the relative shift of $\tc$; solid curve is numeric solution and
dashed curve is the $\sim a\sqrt{|\ln(a)|}$ approximation. 
  }
\end{figure}
We expand $\nbc(\xi_{\vk}) = \nbc(\e_{\vk}-\mu_\c 
+\S_{\rm R}(\vk,\xi_{\vk}))$ around $\e_{\vk}-\mu_\c$: 
\begin{equation}  \label{nonself_expand_f}
\nbc(\xi_{\vk}) 
\approx  \nbc(\e_{\vk}-\mu_\c) + \nbc'(\e_{\vk}-\mu_\c)
\S_{\rm R}(\vk,\e_{\vk}-\mu_\c)) \, \, .
\end{equation}
This provides us with an infrared cutoff, $-\mu_\c$, for our
integrals.  The approximation is non-selfconsistent in the sense that
a modification of the spectrum to a non-quadratic form no longer plays
any role.  Using this expansion, the relative shift in $\tc$ becomes
\begin{equation}  \label{nonself_2integrals}
\frac{\D \tc}{\tc^{(0)}}    ~ \approx ~
-\f{2}{3n}\int \frac{d^3 k}{(2\pi)^3} 
\lbc \nbc(\e_{\vk}-\mu_\c) - \nbc(\e_{\vk}) \rbc 
~-~ \f{2}{3n}\int \frac{d^3 k}{(2\pi)^3} 
\nbc'(\e_{\vk}-\mu_\c) \S_{\rm R}(\vk,\e_{\vk}-\mu_\c)
\end{equation}
Both integrals can be expanded in $|\b_\c\mu_\c|$, as detailed in
appendix \ref{nonself_appendcalc}, each integral being
$\ord(\sqrt{|\b_\c\mu_\c|})$ at leading order.  The result is
\begin{equation}  \label{nonself_result}
\frac{\D \tc}{\tc^{(0)}}   ~ \approx ~
+ \frac{2}{3n} \lbc \f{\sqrt{\pi}}{\l_\c^3}  \sqrt{-\b_\c \mu_\c}\rbc  
+ \ord(\b_\c \mu_\c)
~=~ \f{2}{3} \lbc\zeta(\tfrac{3}{2}) \rbc^{-1} 
\sqrt{\pi|\b_\c\mu_\c|}
+ \ord(\b_\c \mu_\c)
\end{equation}
The infrared cutoff needs to be expressed in terms of the scattering
length; this is achieved by using the condition for the transition,
$\mu_c = \S_{\rm R}(0,0)$.  The relation is
\begin{equation}  \label{nonself_muc}
\b_\c \mu_\c \approx \f{16\pi a^2}{\l_\c^2} \ln(-\b_\c\mu_\c) \, \, ,
\end{equation}
also derived in appendix \ref{nonself_appendcalc}.  The functional
dependence of $|\b_\c\mu_\c|$ on the scattering length is thus similar
to $\sim (a/\l_\c)^2 |\ln(a/\l_\c)|$, and therefore we get a
shift approximately of the form $\Delta\tc \sim a\sqrt{|\ln{a}|}$.  

The approximation of eq \eqref{nonself_muc} by $|\b_\c\mu_\c|
\approx 16\pi(a/\l_\c)^2  |\ln(a/\l_\c)|$ is displayed in figure
\ref{fig_nonself_betamu}.

\section{Self-Consistent Approach: Numerics}  \label{sect_selfcons}

A self-consistent treatment of \eqref{qpdisp} involves a calculation
of the (sub-quadratic) quasiparticle dispersion $\xi(\vk)$.  The
principal values in \eqref{qpdisp} make it difficult to treat
numerically.  A ``high-temperature'' approximation, obtained by using
$f(\xi) \approx 1/\b\xi$, produces a simpler version without the
zeroes of the principal value form: 
\begin{equation}  \label{qpdisp_highT}
\xi_\vk = \e_\vk 
- 2t^2 \int \frac{d^3 p}{(2\pi)^3} \int\frac{d^3 q}{(2\pi)^3} \  
\lba \frac{1} {\xi_{\vp} \xi_{\vq} \xi_{\vp+\vk-\vq} } 
- \frac{1} {\xi_{\vp} \xi_{\vq} \xi_{\vp-\vq} }  \rba  \, \, .
\end{equation}
While the same dimensional arguments apply to \eqref{qpdisp_highT}, it
is likely to lose information about logarithm-like factors.  

Since the self-consistent spectrum has infrared behavior $\xi(\vk)
\sim k^{3/2}$, we look for solutions of the form $\xi(\vk) = k^{3/2}
\s(k) + k^2$ where $\s(x)$ is some smooth function that decays to zero
for large $x$, and is constant or logarithm-like at small $x$.  (Here
the momenta and $\xi$ have been rescaled to be dimensionless, by
factors $\sqrt{\b/2m}$ and $\b$ respectively).  

\begin{figure}
\includegraphics[width=8cm,height=6cm]{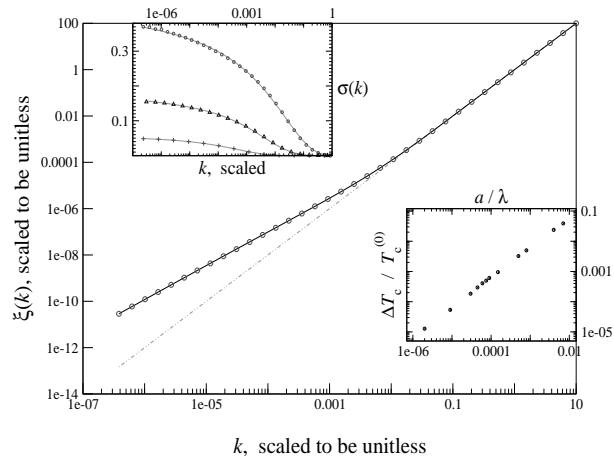}
\caption{   \label{fig_selfcons} 
Quasiparticle spectrum, for $(a/\l) = 0.0005$.  (The ``fit'' indicates
quality of iteration -- line shows penultimate, and dots show final,
iteration).  Crossover from $k^{3/2}$ to $k^2$ behavior is clear;
dotted line is the noninteracting $k^2$ spectrum for comparison.
Upper inset shows the crossover function $\s(k)$ for several $a/\l$
values ($9\times{10}^{-5}$, $8\times{10}^{-4}$ and
$4\times{10}^{-3}$).  Lower inset plots relative shift of $\tc$
against interaction.
 }
\end{figure}

Figure \ref{fig_selfcons} shows a numerically determined spectrum
$\xi(\vk)$.  Once $\xi(\vk)$ has been computed for a particular
$a/\l$, the relative shift in $\tc$ can be computed using our basic eq
\eqref{formula}.  The $\D\tc$ vs. $a$ plot in the inset shows a
definite linear-like increase, as opposed to, say, a $\sqrt{a}$ or
$a^{3/2}$ dependence.  However the existence, or lack thereof, of a
logarithm-like factor is difficult to prove from numeric data.

\section{Concluding Remarks}

To summarize, we have outlined the calculation of $\tc$ from a
perturbative expansion in the two-body $t$-matrix, or equivalently the
scattering length $a$.  The transition temperature is expressed in
terms of the quasiparticle spectrum at $\tc$, and hence
the problem is transformed to one of calculating the critical-point
spectrum $\xi_\vk = E_\vk(\tc)$. 

In developing the formalism, we find that the use of the two-body
$t$-matrix removes the ultraviolet divergences that appear in a
perturbative calculation in terms of a momentum-independent bare
potential $U$.  We also find that the width of the quasiparticle peak
($\S_{\rm I}$) does not affect $\tc$.  Whether this is true at all
orders, or a peculiarity of second-order, and whether this is true for
a more accurate treatment in terms of the many-body $T$-matrix, remain
open questions.

The new feature of this work is an unusual approach to dealing with
the infrared divergence that appears in treating the spectrum at
$\tc$.  This regularization scheme bypasses the issue of spectrum
modification, and results in a non-analytic ($\sqrt{\ln{a}}$-like)
correction factor to the linear shift of the transition temperature.

The major remaining issue in the $\tc$ problem seems to be the
possible existence of a logarithm-like factor in the leading
$a$-dependence.  The resolution of this problem should lie in a clear
treatment of the spectrum at $\tc$.  Of the two major approaches, the
perturbative approach alters the quadratic spectrum too strongly,
while the RG calculation \cite{stoof2} neglects spectrum alteration.

The reason for perturbation theory over-estimating the spectrum
modification remains to be examined in detail.  It has been suggested
\cite{stoof_private} that the reason is that perturbation in $a$ or
$t$ does not take into account the fact that the effective interaction
strength \emph{flows to zero} under renormalization group flow, at the
critical point.  An alternate way to view this is to note that the
many-body $T$-matrix (sum of ladder diagrams) vanishes
\cite{stoof_variational,stoof2,shi+griffin} at the transition
temperature.  A perturbative calculation in terms of the many-body
$T$-matrix, as opposed to one in terms of the two-body $t$-matrix that
we have considered here, may therefore be expected to give a more
realistic modification of the quasiparticle spectrum at $\tc$.  A
calculation of $\xi_\vk$ in terms of the many-body $T$-matrix has not
yet appeared in the literature.

\acknowledgments

The junior author (MH) would like to acknowledge informative
discussions with Henk Stoof and Franck Lalo\"e, made possible by the
generosity of ECT*, Trento (Italy).  E-mail responses from M. Holzmann
and Adriaan M.J. Schakel were very helpful.


\appendix

\section{Details for the Non-SelfConsistent Approach}
\label{nonself_appendcalc}

In this appendix we evaluate the two contributions to
$\D\tc/\tc^{(0)}$ in eq \eqref{nonself_2integrals}, and show how eq
\eqref{nonself_muc} follows from the criticality condition $\S(0,0) =
\mu$.

The first integral in \eqref{nonself_2integrals} is
$\int_\vk\lbc{f}_\c(\e_{\vk}-\mu_\c)-f_\c(\e_{\vk})\rbc =
\l_\c^{-3}g_{3/2}(e^{\b_\c\mu_\c}) - \l_\c^{-3}\zeta(\tfrac{3}{2})$.
Here the $g_{3/2}$'s are the Bose-Einstein integral functions
\cite{BE_functions}, $g_n(x) = \sum_i(x^i/i^n)$, and $\zeta(n) =
g_n(1)$.  We are using the notation $\int_\vk \equiv
(2\pi)^{-3}\int{d^3}k$.  Using the Robinson expansion
\cite{BE_functions} for $g_{3/2}$, we get
\[
\int_\vk \lbc f_\c(\e_\vk-\mu_\c) -f_\c(\e_\vk)
\rbc 
= \f{1}{\l_\c^3} [-2\sqrt{\pi(-\b_\c\mu_\c)} +\ord(\b_\c\mu_\c)]
\, \, .
\]

For the second integral in \eqref{nonself_muc}, using the
second-order self-energy from eq \eqref{SE_real}, one obtains after
some variable transformations
\[
\int_{\vk} f'_\c(\te_\vk) \S_{\rm R}(\vk,\e_\vk) 
= - 2t^2 \f{\del}{\del \mu_\c}\ \int_{\vp}\int_{\vq}\int_{\vk}
\frac{f_\c(\te_\vp) f_\c(\te_\vq) f_\c(\te_\vk)}
{\e_{\vp}+\e_{\vk}-\e_{\vq}  
-\e_{\vp+\vk-\vq}}  \,   
\approx \f{1}{\l_\c^5} 16\pi^{3/2}a^2
\f{\ln(-\b_\c\mu_\c)}{\sqrt{-\b_\c\mu_\c}}   \, \, .
\]
Here $\te_\vp = \e_\vp-\mu_\c$. 
In combination with the criticality condition \eqref{nonself_muc},
this leads to
\[
\int_{\vk} f'_\c(\te_\vk) \S_{\rm R}(\vk,\e_\vk) \approx 
\f{1}{\l_\c^3} \lbc \sqrt{\pi(-\b_\c\mu_\c)} +\ord(\b_\c\mu_\c) \rbc 
\, \, .
\]

The condition \eqref{nonself_muc} for the transition temperature is
obtained from $\mu_\c=\S_{\rm R}(0,0)$ by using the second-order
perturbative result for the self-energy (e.g., obtained from eq
\eqref{SE_real} by setting $\vk = 0$):
\[
\mu_\c= 4t^2\int_{\vp}\int_{\vq}
\frac{f(\te_\vp) f(\te_\vq)}{\e_\vp- \e_\vq -\e_{\vp-\vq}-\mu_\c}
\approx  - \frac{8ma^2}{\pi^2 \b_\c^2} \ln (-\b_\c\mu_\c) \, \, .
\]


\end{document}